\documentclass[epjCONF, onecolumn]{svjour}
\usepackage{graphics}
\usepackage[varg]{txfonts} 
\usepackage[latin1]{inputenc}
\session-title{Conference Title, to be filled}
\begin{document}
\title{Chemical Homogeneity in the Orion Association: Oxygen Abundances of B Stars}
\author{Katia Cunha\inst{1},\inst{2}\fnmsep\thanks{\email{katia@on.br}} \and Ivan Hubeny\inst{2} \and Thierry Lanz \inst{3} }
\institute{Observatorio Nacional - Rio de Janeiro, BR \and University of Arizona - Steward Observatory, USA \and University of Maryland, USA}
\abstract{
We present non-LTE oxygen abundances for a sample
of B stars in the Orion association. 
The abundance calculations included
non-LTE line formation and used fully blanketed non-LTE model atmospheres. 
The stellar parameters were the same
as adopted in the previous study by 
Cunha \& Lambert (1994).
We find that the young Orion stars in this sample of 10 stars are described by a single
oxygen abundance with an average value of A(O)=8.78 and a small dispersion of $\pm$0.05, dex which
is of the order of the uncertainties in the analysis.
This average oxygen abundance compares well with the average oxygen 
abundance obtained previously in 
Cunha \& Lambert (1994): A(O) = 8.72 $\pm$ 0.13 although this earlier
study, based upon non-blanketed model atmospheres in LTE, displayed larger scatter.
Small scatter of chemical abundances in Orion B stars had also been found in
our previous studies for neon 
and argon; 
all based on the same effective temperature scale.
The derived oxygen abundance distribution for the Orion association compares
well with other results for the oxygen abundance in the solar neighborhood.
} 
\maketitle
\section{Introduction}
\label{intro}
Oxygen is one of the most abundant elements and is produced mainly in
hydrostatic burning in massive stars and ejected in the interstellar medium
via Type II SNe. It tracks star formation and  massive
star evolution and is recycled very quickly into the interstellar medium. 
Modelling of the stellar atmospheres of early type stars has advanced in recent years
with improvements in the atomic data making full self-consistent non-LTE calculations possible.
In this contribution we present oxygen abundances for a sample of young stars
in the Orion association.

\section{Analysis and Results}

The target stars are 10 OB main-sequence
stars members of the subgroups of the Orion association.
Full non-LTE calculations were done using non-LTE model atmospheres computed by
TLUSTY \cite{Refc}
with updated and extended model atoms.
The stellar parameters were the same as adopted in the previous study of \cite{Refa}.
Oxygen abundances were obtained from the best fits between observed and
synthetic spectra from a large number of O II transitions.

\label{sec:1}

The average oxygen abundance obtained for the studied sample is: A(O)= 8.78 $\pm$ 0.05 dex.
The scatter in the derived A(O) can be completely explained in terms of the abundance uncertainties
in the analysis.
In Figure 1 we show a comparison of the oxygen abundances obtained here (top panel) with
other recent studies from the literature. 
The middle panel shows results from
\cite{Reff} and \cite{Refg} that were obtained with FASTWIND models. Our
abundances overlap the most recent study from this group but not the lower oxygen from \cite{Reff}.
For comparison, we also show in the bottom panel the results from \cite{Refe}
for a sample of 6 stars in the solar neighborhood: the solar vicinity
abundances and scatter compare well with our results for Orion. Taken together these results indicate
that the chemical abundances in the Orion association are homogeneous and agree well with solar neighborhood values. 
The earlier oxygen abundance distribution from \cite{Refa}
presented more scatter, although with the same average oxygen abundance. 

\begin{figure}
\resizebox{0.55\columnwidth}{!}{%
  \includegraphics {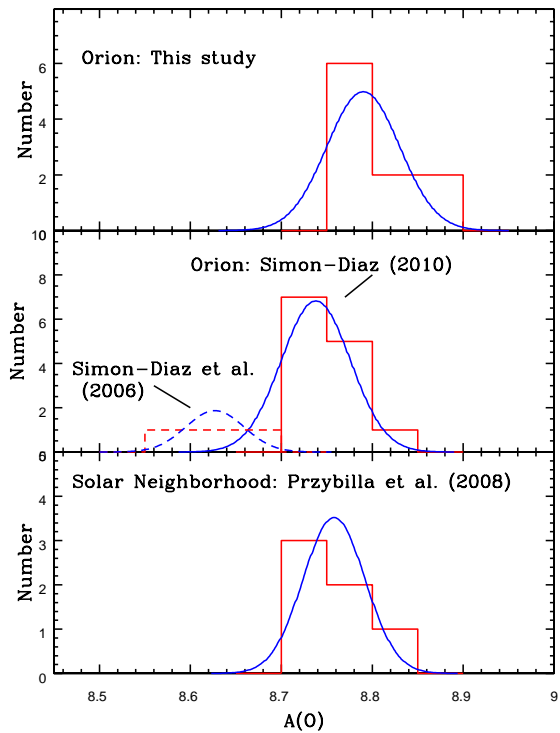} }

\caption{A comparison of our results with other recent results from the literature for
the Orion association (middle panel) and for the solar neighborhood (bottom panel).
The oxygen abundances overall compare well.
Indications that the Orion association chemical composition is homogeneous had been
found in our previous studies for neon 
\cite{Refb}
and argon \cite{Refd}
all based on the same effective temperature scale. Recent results from \cite{Refh}
also confirm this homogeneity.
}
\label{fig:1}       
\end{figure}

\end{document}